\documentclass[twocolumn, trackchanges]{aastex63}
\usepackage{graphicx, amsmath}
\usepackage[caption=false]{subfig}

\usepackage{hyperref}
\usepackage{cleveref}
\usepackage{xspace}
\usepackage{rotating}
\usepackage{comment}

\newcommand{\lya}{Ly$\alpha$\ }
\newcommand{\lyb}{Ly$\beta$\ }

\newcommand{\lyaa}{Ly$\alpha$}

\newcommand{\kms}{km $\mathrm{s}^{-1}$\xspace}
\newcommand{\ergsscma}{ergs~s$^{-1}$~cm$^{-2}$~\AA$^{-1}$\xspace}
\newcommand{\nwmsr}{nW~m$^{-2}$~sr$^{-1}$}
\newcommand{\lfnorm}{$LF_{\mathrm{norm}}$}

\graphicspath{{./}{figures/}}

\begin{document}
\title{Using Lyman Alpha Absorption to Measure the Intensity and Variability of $z \sim 2.4$ Ultraviolet Background Light}

\author[0000-0002-4974-1243]{Laurel H. Weiss}
\affiliation{Department of Astronomy, The University of Texas at Austin, 2515 Speedway Boulevard, Stop C1400, Austin, TX 78712, USA}

\author[0000-0002-8433-8185]{Karl Gebhardt}
\affiliation{Department of Astronomy, The University of Texas at Austin, 2515 Speedway Boulevard, Stop C1400, Austin, TX 78712, USA}

\author[0000-0002-8925-9769]{Dustin Davis}
\affiliation{Department of Astronomy, The University of Texas at Austin, 2515 Speedway Boulevard, Stop C1400, Austin, TX 78712, USA}

\author[0000-0002-2307-0146]{Erin Mentuch Cooper}
\affiliation{Department of Astronomy, The University of Texas at Austin, 2515 Speedway Boulevard, Stop C1400, Austin, TX 78712, USA}

\author[0000-0002-6907-8370]{Maja {Lujan Niemeyer}}
\affiliation{Max-Planck-Institut f\"{u}r Astrophysik, Karl-Schwarzschild-Str. 1, 85741 Garching, Germany}

\author[0000-0001-7066-1240]{Mahdi Qezlou}
\affiliation{Department of Astronomy, The University of Texas at Austin, 2515 Speedway Boulevard, Stop C1400, Austin, TX 78712, USA}

\author[0009-0003-1893-9526]{Mahan Mirza Khanlari}
\affiliation{Department of Astronomy, The University of Texas at Austin, 2515 Speedway Boulevard, Stop C1400, Austin, TX 78712, USA}

\author[0000-0002-1328-0211]{Robin Ciardullo} \affiliation{Department of Astronomy \& Astrophysics, The Pennsylvania State University, University Park, PA 16802, USA} 
\affiliation{Institute for Gravitation and the Cosmos, The Pennsylvania State University, University Park, PA 16802, USA}

\author[0000-0003-2575-0652]{Daniel Farrow}
\affiliation{Centre of Excellence for Data Science,
Artificial Intelligence \& Modelling (DAIM),
University of Hull, Cottingham Road, Hull, HU6 7RX, UK}
\affiliation{E. A. Milne Centre for Astrophysics
University of Hull, Cottingham Road, Hull, HU6 7RX, UK}

\author[0000-0003-1530-8713]{Eric Gawiser}
\affiliation{Department of Physics and Astronomy, Rutgers, the State University of New Jersey, Piscataway, NJ 08854, USA}

\author[0000-0002-5659-4974]{Simon Gazagnes}
\affiliation{Department of Astronomy, The University of Texas at Austin, 2515 Speedway Boulevard, Stop C1400, Austin, TX 78712, USA}

\author[ 0000-0001-6842-2371]{Caryl Gronwall}
\affiliation{Department of Astronomy \& Astrophysics, The Pennsylvania State University, University Park, PA 16802, USA} 
\affiliation{Institute for Gravitation and the Cosmos, The Pennsylvania State University, University Park, PA 16802, USA}

\author[0000-0001-6717-7685]{Gary J. Hill} 
\affiliation{McDonald Observatory, The University of Texas at Austin, 2515 Speedway Boulevard, Stop C1402, Austin, TX 78712, USA}
\affiliation{Department of Astronomy, The University of Texas at Austin, 2515 Speedway Boulevard, Stop C1400, Austin, TX 78712, USA} 

\author[0000-0001-7240-7449]{Donald P. Schneider}
\affiliation{Department of Astronomy \& Astrophysics, The Pennsylvania
State University, University Park, PA 16802, USA}
\affiliation{Institute for Gravitation and the Cosmos, The Pennsylvania State University, University Park, PA 16802, USA}

\begin{abstract}
We present measurements of $z \sim 2.4$ ultraviolet background light using \lya absorption from galaxies at $z \sim 2-3$ in the Hobby-Eberly Telescope Dark Energy Experiment (HETDEX) database. Thanks to the wide area of this survey, we also measure the variability of this light across the sky. The data suggest an asymmetric geometry where integrated ultraviolet light from background galaxies is absorbed by \ion{H}{1} within the halo of a foreground galaxy, in a configuration similar to damped \lya systems. Using stacking analyses of over 400,000 HETDEX LAE spectra, we argue that this background absorption is detectable in our data. We also argue that the absorption signal becomes negative due to HETDEX's sky subtraction procedure. The amount that the absorption is over-subtracted is representative of the $z \sim 2.4$ UV contribution to the overall extragalactic background light (EBL) at \lyaa. Using this method, we determine an average intensity (in $\nu J_{\nu}$ units) of $12.9 \pm 3.7$ \nwmsr~at a median observed wavelength of 4134\,\AA\null, or a rest-frame UV background intensity of $508 \pm 145$ \nwmsr~at $z\sim2.4$. We find that this flux varies significantly depending on the density of galaxies in the field of observation. Our estimates are consistent with direct measurements of the overall EBL.   
\end{abstract}


\section{Introduction}\label{sec:intro}

The extragalactic background light (EBL) is the integrated intensity of light emitted throughout cosmic history across the electromagnetic spectrum. If precisely measured, the spectrum and time evolution of the EBL can be used to constrain models of galaxy formation, galaxy evolution, and the growth of structure.  However, precise measurements are difficult due to contributions from foreground sources such as Zodiacal light, scattered starlight, and scattered Milky Way light \citep{Cooray_2016}. Additionally, the differentiation between the EBL and contributions from local galaxy over-densities is complicated. Any observational measure of the EBL has to account for these issues, and numerical modeling of the background light should similarly include observational effects.

Studies such as \cite{Miralda_Escude_1990}, \cite{Faucher_Giguere_2009}, and \cite{Haardt_Madau_2012} model the radiative transfer of UV emission from active galactic nuclei (AGN) and star-forming galaxies through the inter-galactic medium (IGM), and predict the evolving UV background (UVB) component of the EBL\null. This UVB model is representative of an average measurement over the full sky, whereas most observations of extragalactic background light are localized to a small region of space. In the optical, \cite{Lauer_2022} and \cite{Postman_2024} directly measured the total EBL within a mostly empty 17\farcm 4 region of sky, reporting a cosmic optical background level (COB) of $11.16 \pm 1.65$ \nwmsr~at $\sim 6000$ \AA. This level is higher than that of the predicted integrated galaxy light from deep ground and space-based galaxy counts at similar wavelengths \citep[e.g.][]{Driver_2016, Saldana-Lopez_2021}. The differences between direct flux measurements of the EBL and those derived from indirect methods/modelling are significant and need to be reconciled. One concern in these estimates of extragalactic background light is the effect of cosmic variance and local density enhancements.

EBL measurements over a narrow region of sky will be subject to variations along lines of sight due to over/under-densities of galaxies and other sources. In the limiting case of this effect, contributions to the background would arise from the presence of a single source; a lone quasar may act as the entire ``EBL''. In fact, studies of the \lya forest and damped \lya systems (DLAs) utilize this exact configuration, in which light from a background quasar is used to study structures in the foreground \citep[e.g.][]{Gunn_Peterson_1965, Wolfe_1995, Storrie-Lombardi_2000, Slosar_2011}.

To place constraints on the intensity and on-sky variation of extragalactic background light, an untargeted, wide field survey such as HETDEX (Hobby-Eberly Dark Energy Experiment) is advantageous. In \cite{Weiss_2024} (hereafter referred to as Paper I), we use stacking techniques to demonstrate the detection of faint background light via \lya absorption associated with foreground Lyman-alpha emitters (LAEs). Considering that LAEs at $z \sim 3$ trace the large-scale clustering of galaxies \citep{Gawiser_2007, Guaita_2010, Kusakabe_2018, Ramakrishnan_2024}, our results suggest that the intensity of background light increases in over-dense regions. As a result, observations of these foreground LAEs provide an avenue through which to study the intensity and variation of the EBL at $z \sim 2.5$. This opportunity is unique to HETDEX LAEs due to its large field coverage, sky subtraction procedure, and sheer number of spectra available. 

This paper is organized as follows. Section \ref{sec:data} describes the HETDEX spectra and the selection of data we use.  Section \ref{sec:stacking} discusses the stacking methodology and sky-subtraction. Section \ref{sec:ebl} outlines our method for measuring the EBL and our results. When we provide a measure in kpc, we imply physical units, assuming the Planck 2018 cosmology \citep{Planck_2018} with $\Omega_{\text{m}}$= 0.315 and $H_0$ = 67.4 $\mathrm{km~s^{-1}~Mpc^{-1}}$.

\section{Optical Spectroscopy}
\label{sec:data}

HETDEX \citep{Gebhardt_2021,Hill_2021} is a large, untargeted spectroscopic survey using the upgraded Hobby-Eberly Telescope \citep[HET,][]{Ramsey_1998, Hill_2021}. The survey utilizes the Visible Integral-Field Replicable Unit Spectrograph (VIRUS, \citealt{hil18a, Hill_2021}) that consists of 78 integral field units (IFUs) coupled to 156 spectrographs, with each IFU covering $51\arcsec \times 51\arcsec$ on the sky. Each IFU contains 448 $1\farcs 5$-diameter optical fibers with a 1$/$3 fill-factor, such that a three-position dithered set of exposures provides full spatial coverage within each IFU \citep{Hill_2021}. 
The IFUs are mounted on a 100\arcsec\ grid pattern within a $\approx$18\arcmin\ diameter field of view. 
The optical fibers feed a pair of low-resolution ($750 < R < 950$) spectrographs that cover the wavelength range between 3500\,\AA\ and 5500\,\AA\null. 
The typical exposure time of $\sim 18$ minutes over three dithers then provides 3$\times$34,944 spectra. The final survey area covers $540$ deg$^2$ on sky with a fill-factor of 1/4.6, and corresponding to a co-moving volume of 10.9 Gpc$^3$ over $1.88 < z < 3.52$. 

The HETDEX spectra are sky subtracted and calibrated before being inspected for emission lines and continuum sources as described in \cite{Gebhardt_2021} (we will further discuss the sky subtraction procedure and its significance in this work in Section \ref{sec:stacking}). During inspection, if a source is detected in a fiber, the point-spread-function (PSF) weighted spectrum is extracted from the surrounding fibers. The ELiXer software package \citep{Davis_2023} then classifies the calibrated spectrum and determines the source's redshift.

The LAE spectra in this project come from HETDEX Internal Data Release 4.0.0 (HDR4). This release contains all HETDEX data from January 3, 2017, up to and including August 31, 2023. The updated source catalog contains over 600,000 LAEs with $S/N >5$. For this paper, we select the sources with high-confidence LAE classifications from ELiXer ($P_{\mathrm{Ly}\alpha} > 0.8$) and good quality flags according to the HETDEX catalog (see \cite{Cooper_2023} for the publicly available catalog). We then select objects with where the Gaussian fit to the line yields a $\sigma < 5.5$~\AA\ ($\sim350$ \kms at 4700 \AA) to remove artifacts and/or potential AGN that were not flagged by EliXer or cataloged in \cite{Liu_2022}. This cutoff in line width was determined after visual vetting showed higher artifact contamination at $\sigma > 5.5$~\AA. Our final sample for this paper contains $\sim 400,000$ LAEs across the full HETDEX redshift range.

A significant reduction of noise is necessary to detect the faint EBL flux. This noise reduction is accomplished with spectral stacking, as individual Ly$\alpha$ spectra in HETDEX are not nearly deep enough to detect the EBL contribution. Moreover, while the individual HETDEX spectra are flux calibrated to about 15\% accuracy \citep{Gebhardt_2021}, EBL analyses require calibrations well beyond 1\% accuracy.  Going from 15\% to 1\% requires implementing improvements to baseline HETDEX reductions, specifically regarding sky subtraction.  These procedures are described below. 

\begin{figure*}
    \centering
    \includegraphics[height=5.75cm]{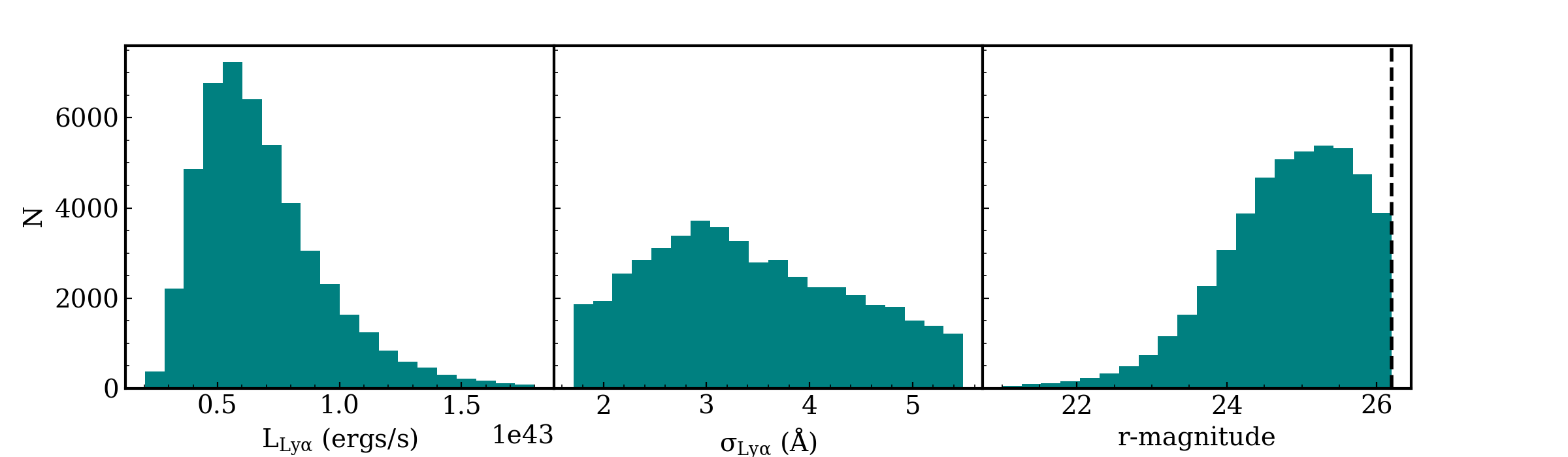}
    \caption{The distribution of a few HETDEX LAE properties (from left to right): \lya line luminosity, linewidth, and counterpart r-magnitude (with a limiting magnitude of $\sim26.2$. Further discussion of the properties of HETDEX LAEs can be found in \cite{Gebhardt_2021}, \cite{Cooper_2023}, and \cite{Davis_2023_hpsc}. In Paper 1, we discussed the effect some of these properties have on the \lya absorption troughs and concluded that the troughs exist across a variety of stacks from a range of properties.} 
    \label{fig:histograms}
\end{figure*}

\subsection{Additional Corrections for Stacking Analyses}
\label{sec:corrections}

Prior to stacking, we perform several corrections to the processed spectra. First, since our goal is to measure the faint EBL, we must be as precise as possible with the sky-subtracting. For each observation, HETDEX measures a global sky using all $\sim 35,000$ fibers distributed over the full 21\arcmin\ diameter focal plane, and a local sky using fibers within the immediate vicinity of each object detection. The latter is based on the signal from 112 fibers which feed an individual CCD amplifier, and covers an on-sky region of $51\arcsec\times 12\farcs5$. We remove the fibers with a continuum detection, i.e. when a fiber contains $> 3\times$ the biweight scale in counts of all fibers on the amplifier. Since we plan to measure the EBL within different fields, we use the local sky-subtraction procedure for this work, which is more precise than the full field sky subtraction. For a more detailed description of HETDEX sky subtraction, see \cite{Gebhardt_2021}. 

Since stacking hundreds to thousands of spectra significantly reduces the noise on our data, we must refine our sky subtraction to a model with an error less than 1\% the value of the sky. On individual spectra, there is no need for this refinement, as the effects of any sky-subtraction residuals only become apparent through the stacking process. To achieve this precision, we statistically construct a representative ``empty'' fiber spectrum (e.g. ``sky only'') on a per-shot (i.e. field) basis to account for residual flux missed by the initial sky-subtraction procedure.  

To create this spectrum for a given observation, we use the sky-subtracted fibers that remain after removing fibers with known issues or significant continuum (from foreground sources or artifacts). These issues include fibers that are located on a bad amplifier, are associated with a meteor or satellite track, have throughput problems, or contain an excessive amount of flagged values. To eliminate fibers containing continuum, we remove all fibers where the average value of the spectrum is $>0.25 \times 10^{-17}$~\ergsscma or $<-0.05\times 10^{-17}$~\ergsscma over the wavelength range 3500\,\AA\ -- 3860\,\AA, or outside $\pm 0.05\times10^{-17}$~\ergsscma in any of following bandpasses: 3860--4270\,\AA, 4270--4860\,\AA, 4860--5090\,\AA, and 5090--5500\,\AA\null. These fluxes and ranges were chosen via calibration to SDSS g-band magnitudes. Lastly, for the remaining fibers, we remove the top 1\% of fluxes in each wavelength bin to further ensure exclusion of the continuum just below the cuts.  We then stack the data to create a single ``empty'' weighted biweight spectrum for a shot. This residual spectrum is subtracted from each fiber spectrum associated with the LAEs in the corresponding shot. 

After applying this correction, we shift the spectra to rest frame wavelengths, since the method of measuring the EBL requires that all spectra be aligned at \lyaa. We convert the observed air wavelengths to vacuum via \cite{Greisen_2006} before shifting to the rest frame using redshifts determined by HETDEX \citep{Cooper_2023}. Because the EBL (by definition) is measured in the observed frame, we only correct the wavelengths for redshift and do not account for any other effects (such as cosmological dimming, Milky Way dust extinction, etc.). Since we are stacking, the EBL estimates will be measured at a mean observed frame wavelength for a given redshift bin.  

\subsection{Stacking Methodology}

Stacking hundreds to thousands of HETDEX LAE spectra increases not only the $S/N$ of the \lya emission line but also the $S/N$ of other features within the observation. The increase in $S/N$ is roughly proportional to the square root of the number of sources. As a result, stacks of $\sim$1000 spectra increase the $S/N$ of a contributing source by a factor of $\sim$30. This facilitates the detection of any faint background signal that is not removed by sky subtraction. Additionally, since HETDEX is an untargeted survey, large stacks of LAEs include different instrument orientations, environment, geometries, and lines of sight. As a result, we have the unique opportunity to investigate the intensity and variation of the global EBL.

We use the stacking method described in \cite{Davis_2021, Davis_2023_hpsc} and Paper I. Briefly, the extent of the restframe wavelength coverage is determined by the objects with the highest and lowest redshifts in the stack, with the grid spacing adopted from the highest redshift object ($0.44$\,\AA\ for $z=3.5$). We then linearly interpolate all the restframe spectra onto the adopted grid and stack each wavelength bin using a weighted biweight statistic---a modified version of Tukey's biweight estimator \citep{Andrews_1972, Beers_1990} where the spectral points are weighted by the inverse of the flux variance \citep{Davis_2021}. The choice of statistic is largely inconsequential, as stacks of $\sim$10,000 or more spectra show little difference when using the mean, median, biweight, or weighted biweight statistics \citep{Davis_2023_hpsc}. For this analysis, stacks of $\gtrsim$1000 spectra using different stacking statistics show little difference in overall absorption depth.

For all spectra used in this paper, we select from a sample of $\sim 300,000$ high confidence LAEs as described in Paper 1. The distribution of \lya luminosities, linewidths, and r-magnitudes are shown in Figure \ref{fig:histograms}. Note that we are sensitive to $r\sim26.2$ and not every HETDEX source has a measured counterpart magnitude. For a discussion of how HETDEX LAE properties affect the absorption troughs, see \S4.3 of Paper 1.

\section{Spectral Stacking}
\label{sec:stacking}

\begin{figure*}
    \centering
    \includegraphics[height=6.0cm]{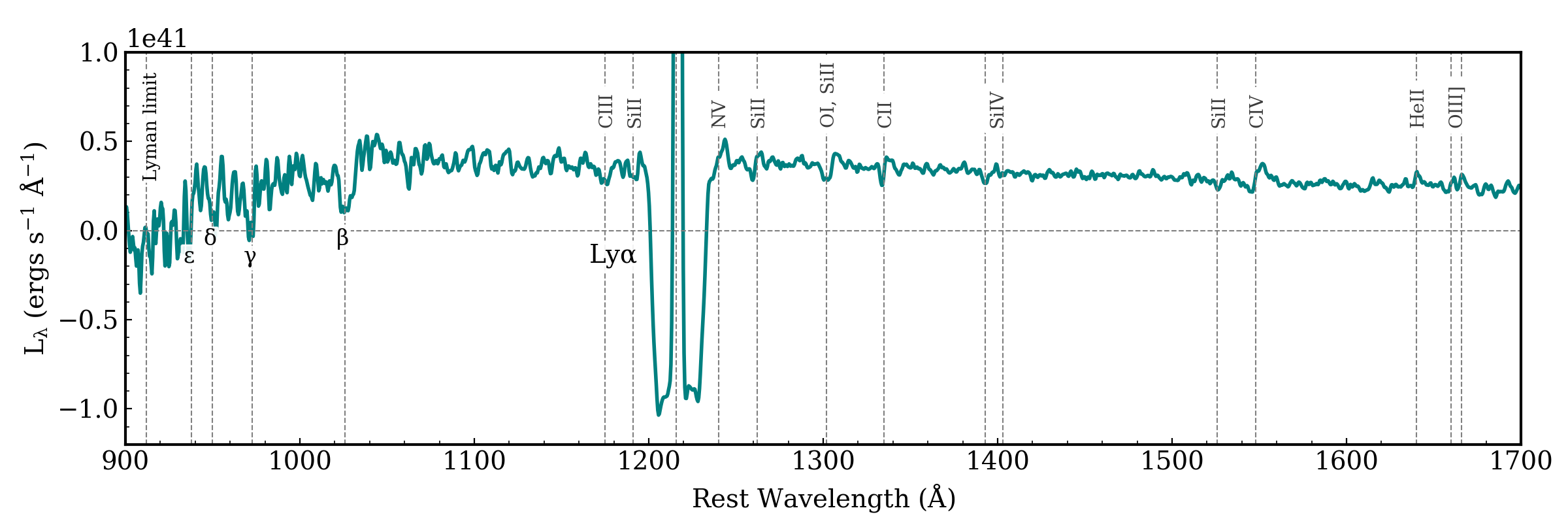}
    \caption{A stack of $\sim 50,000$ high-confidence LAE spectra from HDR4 with $S/N > 5$, similar to the stack presented in \cite{Davis_2023_hpsc}. We select the spectra that have \lya line width $\sigma < 5.5$ \AA~to eliminate unidentified AGN contaminants. The significantly negative flux values of the \lya absorption troughs are likely the result of background over-subtraction as discussed in Paper I.}
    \label{fig:50k stack}
\end{figure*}

\subsection{Effect of Sky Subtraction}

Figure \ref{fig:50k stack} depicts a stack of the roughly 50,000 high-confidence LAEs shown in Paper I and \cite{Davis_2023_hpsc}. As discussed at length in Paper I, the significant negative absorption troughs red- and blue-ward of the \lya emission line are are not purely the result of instrumental or algorithmic effects, but have a real, physical component. This absorption is unique to stacks of LAEs, as they are not present in similar stacks of HETDEX [\ion{O}{2}] emitters, which are detected as frequently and calibrated the same way as LAEs. They are only classified as as a \lya or \ion{O}{2} post processing. These troughs do not appear in stacks of empty (random) fibers and are present in stacks of various LAE subsamples.

\begin{figure}[t]
    \centering
    \includegraphics[height=6cm]{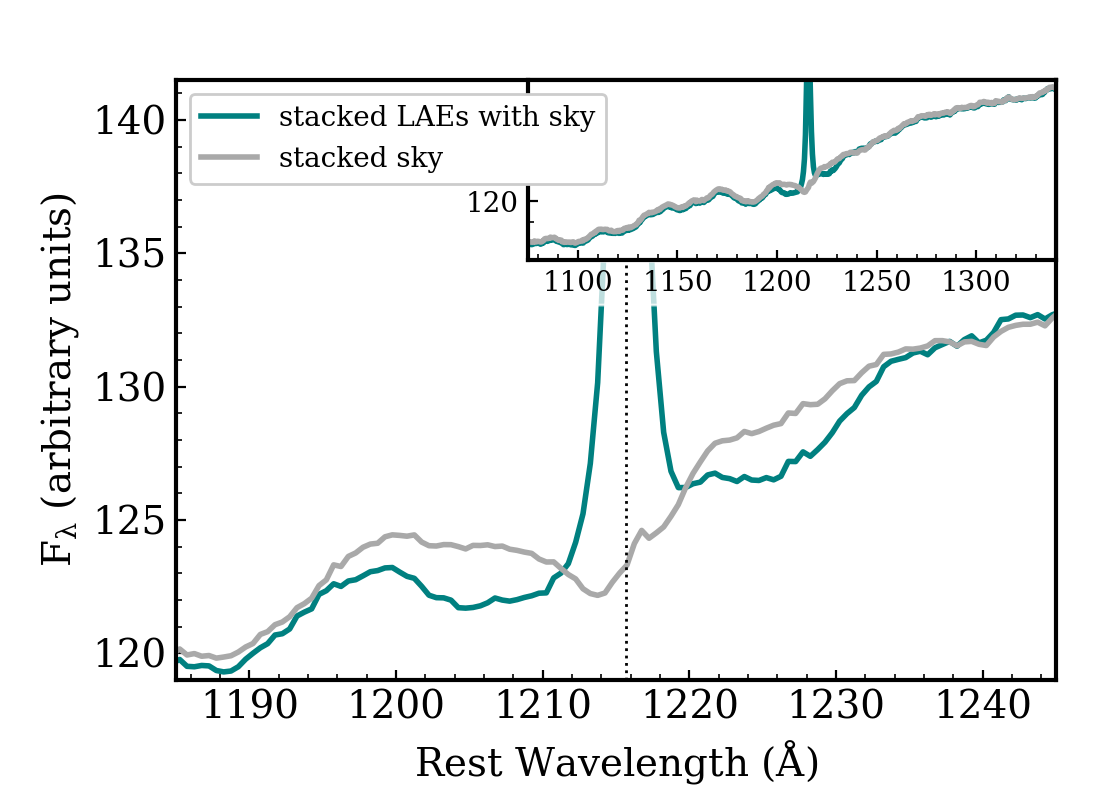}
    \caption{\textit{Main panel:} A zoomed-in plot of the mean stacked fiber spectrum of $\sim$70,000 high-confidence LAEs prior to the sky subtraction and flux calibration procedure (teal line). The $y$-axis is plotted in arbitrary units.  There is a clear deficit of flux-density immediately surrounding the \lya line.  A stack of the corresponding sky spectra calculated for each fiber is plotted in gray. The sky spectrum (gray line), fit to the full amplifier where the corresponding LAE is located, closely matches the stack of LAEs except in the region around \lyaa. While the sky does display some absorption centered on \lya (perhaps due to background absorption from neutral hydrogen not associated with the LAEs), the broad absorption which is easily seen in the LAE stack is not present. In fact, while nearly identical at most wavelengths, the stacked sky spectrum has \textit{more} flux  than the pre-sky-subtracted LAE stack on either side of the \lya emission. \textit{Inset panel:} A zoomed-out plot of the stack of LAE spectra prior to sky subtraction (teal) and the stacked sky spectrum (gray).}
    \label{fig:with sky}
\end{figure}

To examine the possible influence of our sky-subtraction procedure on the troughs, we compare stacked LAE spectra to stacked ``empty-sky'' fiber spectra prior to sky-subtraction. For a subset of $\sim$70,000 high-confidence LAE detections, we select the nearest fiber to each detection's coordinates and save the fiber's uncalibrated spectrum before sky subtraction, in units of photon counts, and the sky spectrum applicable to that fiber. The sky spectrum is a single spectrum modeled across the corresponding amplifier for the detection and then normalized to the selected fiber. While it is possible that some extended \lya emission is picked up in the measurement of ``sky'', the on-sky size of an amplifier in which the sky is measured spans far beyond the typical extent of a \lya halo. We shift each spectrum to the rest-frame wavelength of the LAE and calculate the mean stack, the pre-sky-subtracted spectra, and the sky spectra on a common rest-frame wavelength grid. These stacks are depicted in Figure \ref{fig:with sky}. We find that the troughs exist in the LAE spectra before sky subtraction, though they are, by construction, non-negative. Notably, the mean spectrum of LAEs prior to sky subtraction is lower around the \lya line than the surrounding mean sky spectrum; the presence of an LAE seems to remove some of the sky flux. On average, when the sky is subtracted from an LAE spectrum, the troughs around the \lya emission line become negative. This result supports the background light absorption scenario proposed in Paper I and outlined below.

\section{Measuring the Background Light}
\label{sec:ebl}

For the remainder of this paper, we will use the terms ``background light'', ``UVB'', and ``EBL'' interchangeably. In the rest frame of the LAE, the background light that gets absorbed is in the ultraviolet, although we measure it in the optical. This observed-frame flux measurement more closely aligns with definitions of the EBL (or COB when referring specifically to optical wavelengths). That said, the UVB at high redshifts is, by definition, a component of the EBL, since the integrated UV flux from $z = \infty$ shifts into longer wavelength regimes in the observed frame. For clarity, we will note in our comparisons to other studies how our definition of UVB/EBL differs from other measurements. 

If we assume the physical model of the absorption troughs presented in Paper I, we can use the amount that the troughs are over-subtracted to estimate the level of extragalactic background light experienced by the LAE\null.  We summarize the model and its relevance in measuring the UVB/EBL below. 

\subsection{Physical Model of \lya Absorption Troughs}

In Paper I, we suggested a scenario that explains the existence of negative flux values associated with the \lya absorption troughs shown in Figure \ref{fig:50k stack}. In this model, \ion{H}{1} gas in and around an LAE absorbs diffuse background light at \lya, in a geometry similar to that of DLAs. Since the LAE also emits \lyaa, the result is a combined profile of a \lya emission line sitting within a broad absorption well. In summary, a HETDEX observation of an LAE prior to sky subtraction contains: 

\begin{equation}
    UVB + LAE - UVB_{\mathrm{Ly}\alpha} + sky_f
\label{eq1}
\end{equation}

Where the UVB is the contribution from $z=\infty$ to $z=z_{\mathrm{LAE}}$, $UVB_{\mathrm{Ly}\alpha}$ is the UVB around the \lya transition, and the ``foreground sky'' ($sky_f$) is the contribution from $z=z_{\mathrm{LAE}}$ to $z=0$. In HETDEX processing, the ``sky'', which is measured off-source, cannot distinguish between foreground and background light, thus contains: 
\begin{equation}
    UVB + sky_f
\label{eq2}
\end{equation}
During sky subtraction, \ref{eq2} is subtracted from \ref{eq1} and the expression becomes: 
\begin{multline}
    UVB + LAE - UVB_{\mathrm{Ly}\alpha} + sky_f - (UVB + sky_f) \\
    = LAE + (-UVB_{\mathrm{Ly}\alpha})
\label{eq3}
\end{multline}
Since there is little to no stellar continuum detected in HETDEX LAEs, the oversubtracted UVB around \lya becomes  negative in the overall spectrum. A graphic outlining this process is shown in Figure 7 of Paper I.

\begin{figure}[t]
    \centering
    \includegraphics[height=6cm]{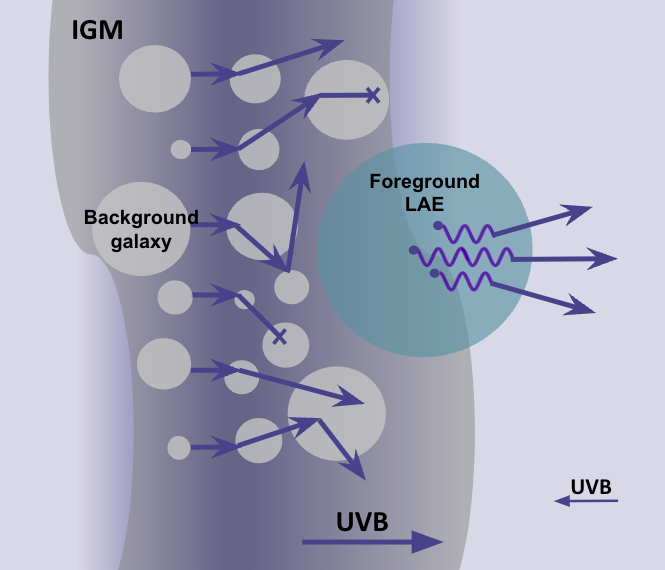}
    \caption{A graphic depicting how an overdensity of background galaxies creates anisotropy in the UVB\null. The background galaxies are depicted as gray circles, with their UV emission indicated by purple arrows that scatter through the intervening IGM (gray band). The purple gradient indicates the strength of the UVB\null. The foreground LAE (teal) produces \lya photons (purple wavy arrows) that are able to reach an observer. From the perspective of an observer located on the right, an LAE sitting on/near the edge of an overdense region experiences a stronger UVB from the increased number of background galaxies. LAEs that exist in a configuration such as this are more likely to be detected by untargeted surveys, as less \lya flux is likely to escape configurations with more galaxies/IGM between the observer and LAE\null. In this geometry, the anisotropy of the UVB is a direct result of the location bias that exists for the most easily detectable LAEs.}
    \label{fig:detailed graphic}
\end{figure}

We suggested in Paper I that HETDEX LAEs likely tend to reside with fewer intervening sources between the LAE and the observer, with more galaxies in the background of an LAE than in the foreground. The resulting UVB experienced by a typical LAE is then anisotropic, which suggests that the gas and dust in and around the LAE is also asymmetric.  Consequently, from the perspective of an observer, more UVB photons at \lya are scattered out of the line of sight than into the line of sight, creating the observed absorption well. Figure \ref{fig:detailed graphic} depicts a simplified graphic of this configuration. 

\begin{figure*}[t]
    \centering
    \includegraphics[height=6cm]{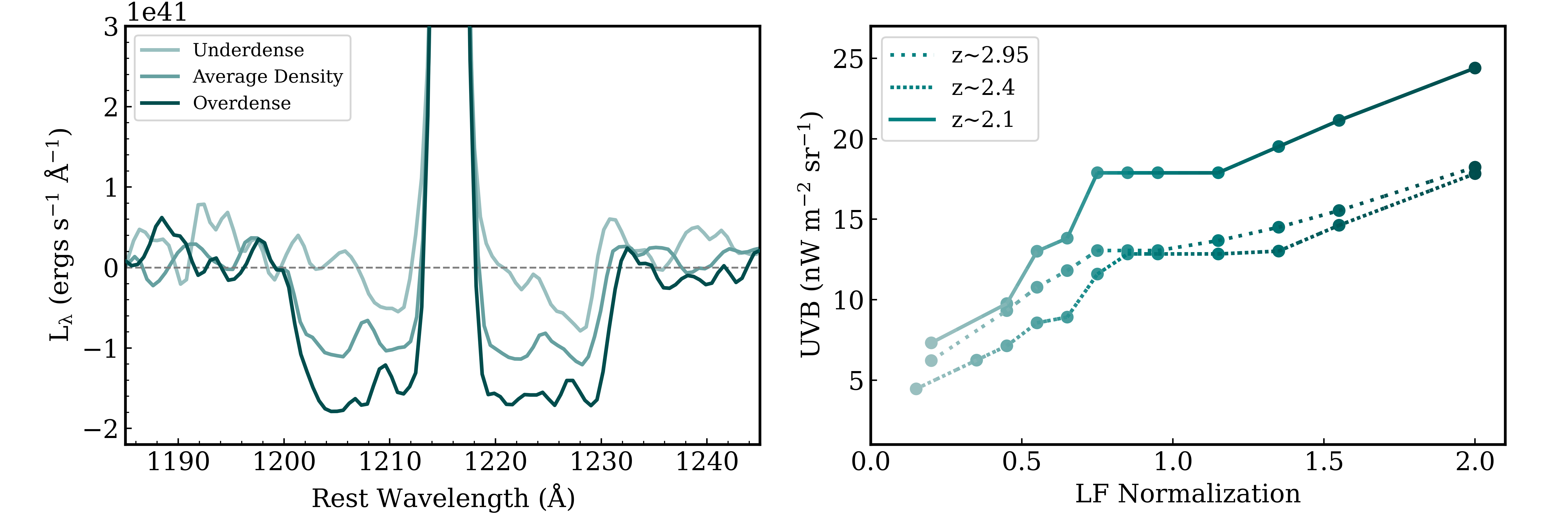}
    \caption{\textit{Left panel:} Stacks of LAEs in underdense, average density, and overdense regions using the field luminosity function normalizations also presented in Paper I. Each stack contains between $\sim$70 and 200 contributing spectra and are restricted to galaxies with $2.3 < z < 2.5$. According to our model, the depth of the absorption troughs increases with field density due to the increasing strength of the UVB in these regions. \textit{Right panel:} Estimates of the EBL intensity (i.e. the UVB measured in observed-frame) determined via the depths of the troughs as a function of luminosity function normalization within three redshift/wavelength bins. Our method for measuring the UVB/EBL is biased against low-density regions (i.e. if there is no LAE to absorb the background, we cannot detect the UVB), thus the measurement is incomplete at low LF normalization values. In addition to the increase in UVB/EBL strength with field density, there may be slight evolution with redshift, though further analysis is needed. }
    \label{fig:stacks v lf, uvb v lf}
\end{figure*}

One intriguing consequence of this model is that the level of the UVB seen by a $z\sim2.4$ LAE is encoded within the absorption troughs. Specifically, the \textit{amount the troughs are over-subtracted in the HETDEX spectra} reflects the UVB absorbed by \ion{H}{1} in the halo of LAEs. Our method of measuring the UVB/EBL is simple: we determine the flux offset in the observed frame that must be added to each individual spectrum in order to make the \lya troughs in the overall stack non-negative. This offset is the ``EBL'' that is over-subtracted in the optical. Shifting this flux level to the rest frame of the LAE, the offset becomes a measurement of the rest frame UVB\null. Put simply, we can use the negative depth of the troughs to measure the UVB at $2 \lesssim z \lesssim 3$, with minimal assumptions.

We first assume that the absorption of the UVB by an LAE is saturated, as in a DLA\null. We note that we have not fully quantified the properties and physical extent of the absorbing gas; this simplifying assumption is based on the apparent shape of the absorption. Additionally, since our methodology of shifting the troughs to zero does not account for the presence of any underlying stellar continuum (see Figure \ref{fig:50k stack} for an example), we assume that the continuum is undetectable in our co-added LAE spectra.  In sufficiently large stacks and stacks of galaxies with bright counterparts, this assumption results in a lower limit to the UVB measurement.  While, in theory, we could shift the troughs to an estimated continuum level near \lyaa, we would also have to assume there is no stellar continuum absorption at \lyaa. However, since the assumption of no detected stellar continuum is consistent with the majority of our stacks of HETDEX LAEs (see the left panel of Figure \ref{fig:stacks v lf, uvb v lf}, which contains the stacks we use for our measurement of the UVB/EBL), we simply shift the troughs to 0. In reference to the detectable continuum in Figure \ref{fig:50k stack} (which we note contains a significant number of spectra) shifting the troughs to 0 requires an additive offset of $+0.063 \times 10^{-17}$ \ergsscma, while shifting the troughs to the continuum level requires an offset of $+0.084 \times 10^{-17}$ \ergsscma. As we will show in the next section, the effect that LAE environment has on the offset is much more significant.

 
\subsection{Effect of Local Density Enhancements}

As presented in Paper I, the strength of the absorption troughs increases with increasing field density. As a proxy for regions of over- and under-density, we simply use the total number of LAEs in a particular field compared to expectations. We do this by measuring the luminosity function (LF) for each HETDEX field, and the overall normalization of the luminosity function (i.e. the number of observed LAEs divided by the number of expected LAEs) becomes the surrogate for overdensity integrated over the redshift range of HETDEX. This work will be presented in Gebhardt et al.\  (in preparation). 

We then divide a sample of high-confidence LAEs at $z\sim$ 2.4 into whether they exist in an underdense ($0 <$ \lfnorm $< 0.3$), average density ($0.95 <$ \lfnorm $< 1$), or overdense ($1.5 <$ \lfnorm $< 1.6$) region of space. Note that the bin widths and limits reflect significantly distinct environments while ensuring a sufficient number of galaxies that falls into each bin. The left panel of Figure \ref{fig:stacks v lf, uvb v lf} depicts the stacked spectra for each overdensity bin. The troughs are strongest in the stack of LAEs that reside in overdense regions, and almost disappear in the stack of LAEs in the most underdense regions. This trend supports our UVB absorption scenario, since overdense fields are likely to contain significantly more background light than fields with a single isolated LAE\null. This supports our interpretation that background light absorption by neutral hydrogen near/around LAEs gives rise to the \lya absorption troughs. 

Using these effects and the methodology described in the previous section, we can measure the intensity of the UVB experienced by a typical LAE at $z\sim$ 2 as a function of field density. While often assumed to be roughly isotropic \citep[e.g.,][]{Faucher_Giguere_2009, Haardt_Madau_2012}, the intensity of UV radiation from background sources (i.e., the UVB) can vary across small scales, due to the anisotropic density of star-forming galaxies and AGN across the sky. Since LAEs detected by HETDEX are biased towards the configuration depicted in Figure~\ref{fig:detailed graphic} (fewer galaxies in the foreground than in the background), this variation is observable via \lya absorption, as the asymmetry causes more photons scatter out of than into our line of sight. As a result, the small-scale density variation and observational bias of HETDEX LAEs creates a unique opportunity through which we can measure the intensity and variability of the UVB. 

\begin{figure*}[t]
    \centering
    \includegraphics[height=6cm]{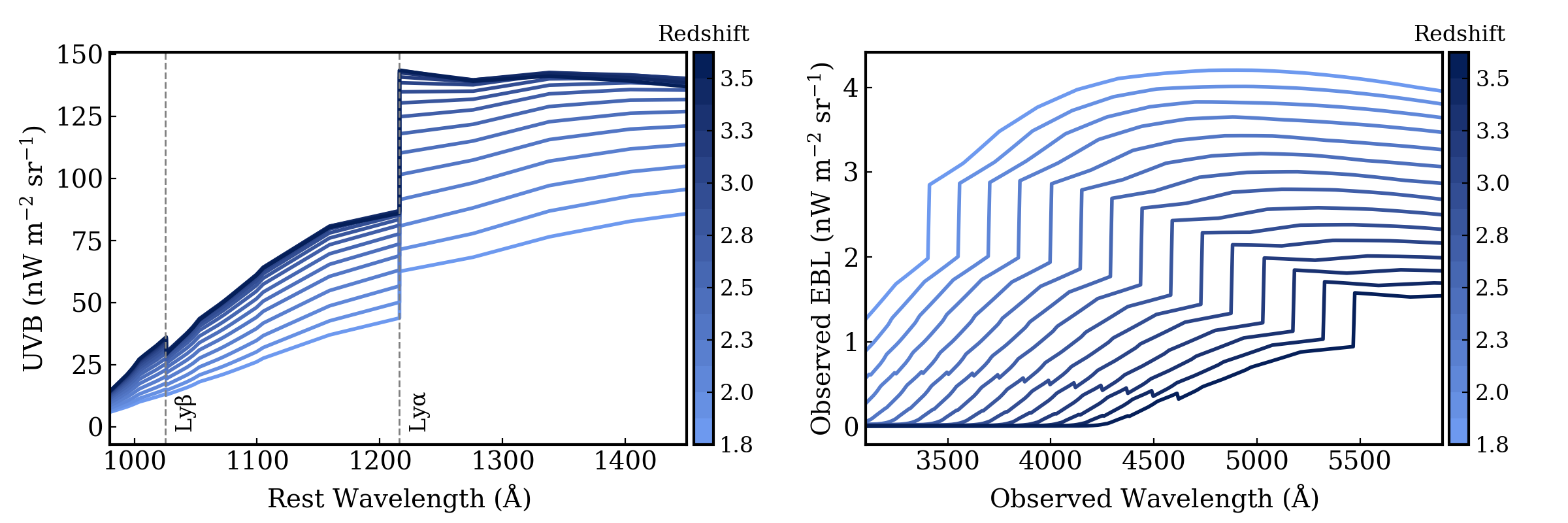}
    \caption{\textit{Left panel:} The rest-frame UVB as modeled by \cite{Haardt_Madau_2012} over the range of redshifts observed by HETDEX\null. Each spectrum is representative of the integrated light from galaxies and AGN from $z = \infty$ to a given $z$ propagated through an evolving IGM\null. The sharp features at \lya and \lyb are the result of Lyman series resonant absorption due to cosmic hydrogen and helium in the IGM\null. For a full description of the radiative transfer methods used to produce these spectra, see \cite{Haardt_Madau_2012}. \textit{Right panel:} The same UVB spectra shifted to observed frame using the appropriate $(1+z)^4$ surface brightness dimming correction. These spectra reflect the contribution from the UVB to the overall EBL in the observed frame. Note: this visualization neglects the contributions from a given $z$ to $z=0$, which are included in conventional definitions of the EBL. }
    \label{fig:uvb}
\end{figure*}

The right panel of Figure \ref{fig:stacks v lf, uvb v lf} plots our estimates of the EBL (or more precisely, the $z \sim 2-3$ UVB intensity shifted to observed-frame) measured using the amount that the troughs are over-subtracted as a function of the luminosity function normalization. Since the troughs occur at \lya in the rest-frame of the LAEs, we can effectively measure the level of the EBL at different wavelengths. The rest-frame UVB at different redshifts should be reflected in the observed-frame EBL as a function of wavelength (we again note that our translation of rest-frame UVB to observed-frame EBL neglects the integrated contribution of light between $z\sim 2-3$ and $z = 0$). By stacking LAEs in three redshift bins, we can obtain a rough estimate of the EBL at three wavelength points. 

HETDEX measurements are taken using an aperture with an effective area of $\sim 9.85''$ (based on average seeing), and their cataloged spectra are in units of \ergsscma\. To compute the UVB/EBL intensities for each overdensity and redshift bin, we convert the bin's observed frame additive offset into surface brightness density units via the unit conversion:
\begin{multline}
    \frac{\mathrm{nW}}{\mathrm{m^2~sr}} = 
    \frac{\mathrm{ergs/s}}{\mathrm{cm^2~\AA~aperture}} \times \frac{1~\mathrm{aperture}}{9.85''} \times \\ \frac{4.25\times10^{10}~\mathrm{''}}{\mathrm{sr}} \times \frac{100~\mathrm{nW}}{\mathrm{ergs/s}} \times \frac{10^5~\mathrm{cm^{2}}}{\mathrm{m^2}} \times \lambda_{\mathrm{Ly\alpha,obs}}~\mathrm{\AA}
\end{multline}

where $\lambda_{\mathrm{Ly\alpha,obs}}$ is the observed frame \lya wavelength for the redshift bin. The resulting surface brightness intensities are shifted to rest frame using the appropriate $(1+z)^4$ surface dimming correction to reflect the rest frame ``UVB'' experienced by an LAE.  

As shown in the right panel of Figure \ref{fig:stacks v lf, uvb v lf}, our measurements of the EBL vary slightly with redshift/observed wavelength, while the trend with LF normalization/field density is much more apparent. These trends in EBL/UVB intensity scale roughly linearly with LF normalization, which is a reasonable outcome of our physical model for the absorption troughs. Further analysis using luminosity functions that vary with redshift and environment are needed to more accurately quantify the variation of the UVB/EBL due to the effect of both local density enhancements and redshift. 

To calculate an average value of our EBL measurements, we weight each value in Figure \ref{fig:stacks v lf, uvb v lf} for $z\sim$ 2.4 by the number of fields in the corresponding luminosity function normalization bin and take a weighted average. This step effectively weights each estimate by the area of sky in which it was measured for a more representative average EBL intensity. We determine an EBL estimate and $1\sigma$ uncertainty of $12.9 \pm 3.7$ \nwmsr~evaluated at a median wavelength of 4134\,\AA and the typical effective area of the HETDEX PSF. This measurement corresponds to a rest-frame UVB intensity of $508 \pm 145$ \nwmsr at $z\sim 2.4$. 

\section{Discussion}

To place our measurements of the UVB/EBL in the proper context, we compare our estimates to a theoretical simulation of the UVB and direct observational measurements of the EBL\null. \cite{Haardt_Madau_2012} generate an evolving spectrum of the UVB by modelling the radiative transfer of UV emission from galaxies and AGN through a clumpy IGM\null. To compare to our observed frame optical measurements, we redshift these UVB spectra using the appropriate $(1+z)^{4}$ surface brightness dimming correction.  Figure \ref{fig:uvb} depicts these rest (left-panel) and observed (right-panel) frame spectra over the range of redshifts observed by HETDEX\null. At the wavelength of \lyaa, the observed frame intensity of the \cite{Haardt_Madau_2012} UVB is roughly $2-3$ \nwmsr~within the HETDEX redshift range. By comparison, our estimates of the UVB span $\sim 4-25$ \nwmsr~depending on field density and redshift. While these measurements agree within an order of magnitude of each other, we stress that our measurements are sensitive to local density enhancements, while \cite{Haardt_Madau_2012} assume an isotropic UVB\null. It is also important to note that the exact shape of the locally enhanced UVB we observe via stacking HETDEX LAEs is likely not identical to the shape of the UVB modeled by \citep{Haardt_Madau_2012}.

\begin{figure}
    \centering
    \includegraphics[height=6cm]{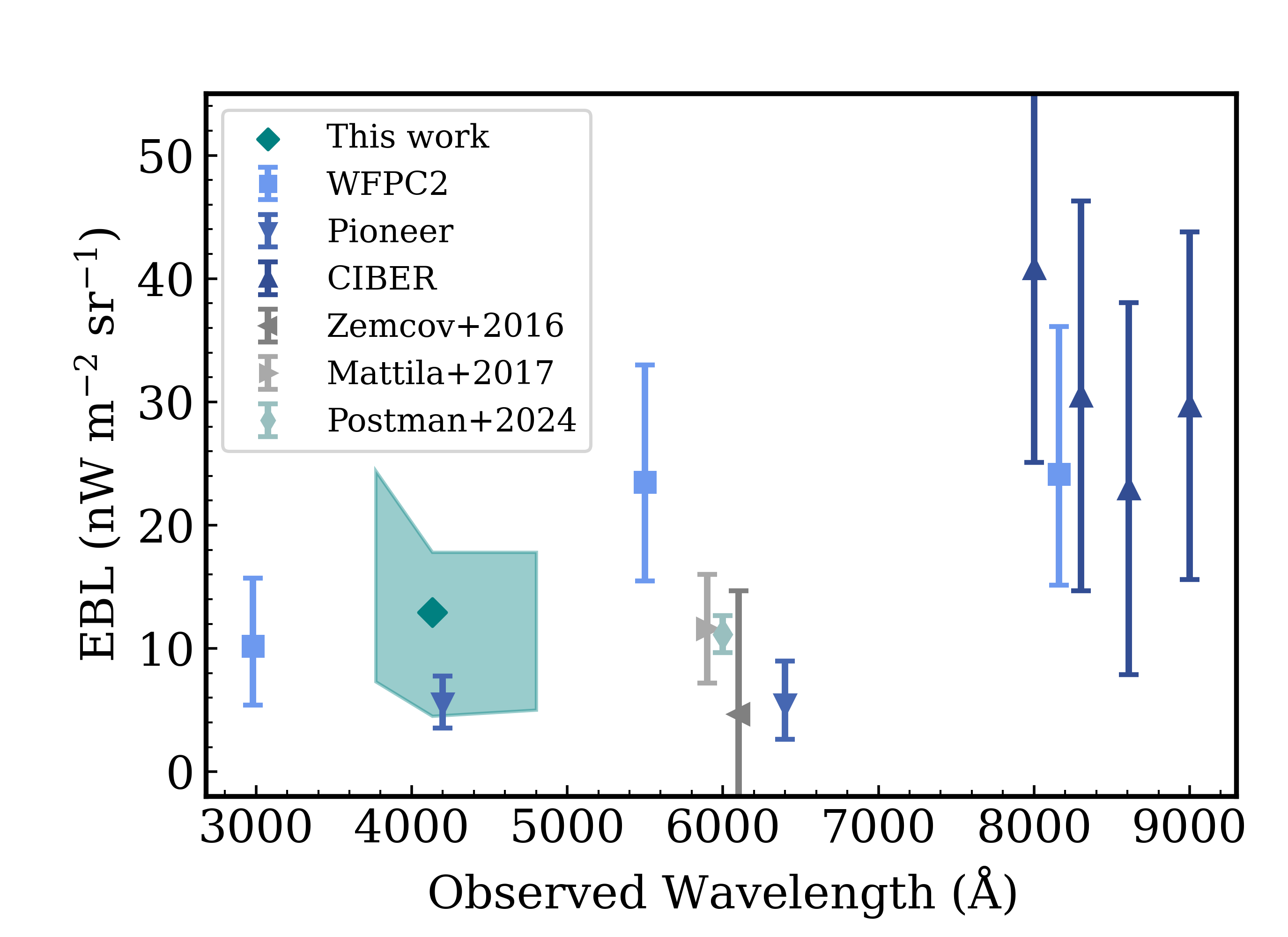}
    \caption{Our measurement of background light compared to observations of the optical EBL (also known as the COB) from WFPC2 \citep{Bernstein_2007}, Pioneer \citep{Matsuoka_2011}, CIBER \citep{Matsuura_2017}, \cite{Zemcov_2017}, \cite{Mattila_2017}, and \cite{Postman_2024}. (The latter three measurements are all at 6000~\AA; they are displayed at slightly different wavelengths for viewing purposes.) Although these studies measure the EBL/COB via different methods and at different wavelengths, there is rough agreement (within an order of magnitude) between the studies. Our ``EBL'' measurement is consistent with these measurements, though we note our value only does not include any contribution from the given $z$ to $z = 0$. The shaded region around our EBL measurement spans the range of values shown in the right panel of Figure \ref{fig:stacks v lf, uvb v lf}. }
    \label{fig:ebl studies}
\end{figure}

Direct observational measurements of the EBL are difficult. They must account for foreground components within the Solar system (such as Zodiacal light in the optical and infrared), as well as Galactic emission from the Milky Way in radio, infrared, X-ray and gamma ray wavelengths \citep{Cooray_2016}. Measurements of the EBL in the optical, or COB, are mainly limited by the choice of ``empty'' regions of sky, as well as the modelling and removal of Zodiacal light and Gegenschein. In the work of \cite{Lauer_2022} and \cite{Postman_2024}, which uses imaging from NASA's New Horizons spacecraft to measure the COB within a high-galactic-latitude field (where the effects of scattered Zodical and Milky Way light are minimal), they still find disagreement with the COB intensity implied by galaxy counts. Figure \ref{fig:ebl studies} plots several studies of direct COB measurements as well as the range of EBL values we estimate here. Since our indirect measurements of the ``optical EBL'' (the redshifted $z \sim 2$ UVB contribution) neglects contributions from $z \sim 2$ to $z=0$, we can interpret this value as a lower limit, which falls within estimates from other COB studies.  We can then use \cite{Haardt_Madau_2012} to estimate the relative contribution between $0<z<2$ and $2<z<\infty$.  According to their model, about 60\% of EBL should be coming from $2<z<\infty$ sources. Thus, in order to compare to the other COB studies, one would need to increase our UVB values by about 67\% to account for the foreground contributions. Since our average value estimate is likely biased towards higher density regions,  our lower bound on the EBL more closely aligns with the methodology and results of other studies. Within the uncertainties, all the measurements are consistent with each other.

We also show that the intensity of the EBL (or UVB, or COB) is very sensitive to the region of sky in which it is measured, due to the non-isotropic distribution of galaxies. More careful interpretation of EBL levels in the context of cosmic variance may reconcile the tension between the measurements produced by theoretical, direct, and indirect methods. 

\section{Summary}

We have presented an indirect method for measuring the EBL from $2 \lesssim z \lesssim 3$ LAEs via \lya absorption of UV background light. We measure the EBL as a function of local density, with a range of 5--18 \nwmsr at $\lambda = 4134$\,\AA, with an average value of $12.9\pm3.7$ \nwmsr. Our measurements find rough agreement with both direct observations of the EBL in the optical, and simulations of the evolving UVB\null. We also show that the intensity of extragalactic background light, when measured in this way, is highly dependent on local density enhancements in the environments of LAEs. This variation emphasizes that the UVB/EBL is likely not isotropic and may vary by as much as a factor of 10 depending on the density of sources in the field and along the line of sight.  

We note that our measurements of the EBL rest on the assumption that our physical interpretation of the \lya absorption troughs is correct. While in Paper I we showed that these troughs are \textit{not} the result of algorithmic and/or instrumental effects, our physical model of the absorption is not complete. Further work is needed to characterize the absorbing gas using radiative transfer, to better quantify the effect of density enhancements, and to investigate redshift evolution. 

Measuring the local EBL is complicated for a variety of reasons, and multiple avenues are needed to obtain a reliable value. The additional leverage provided by the saturated \lyaa\ absorption feature potentially allows for a robust measure to the $z \gtrsim 2$ component of this light.  With its large sky coverage, HETDEX is primed to exploit this measure, and measure its dependence on both redshift and environment.

\acknowledgments

HETDEX is led by the University of Texas at Austin McDonald Observatory and Department of Astronomy with participation from the Ludwig-Maximilians-Universit\"at M\"unchen, Max-Planck-Institut f\"ur Extraterrestrische Physik (MPE), Leibniz-Institut f\"ur Astrophysik Potsdam (AIP), Texas A\&M University, The Pennsylvania State University, Institut f\"ur Astrophysik G\"ottingen, The University of Oxford, Max-Planck-Institut f\"ur Astrophysik (MPA), The University of Tokyo, and Missouri University of Science and Technology. In addition to Institutional support, HETDEX is funded by the National Science Foundation (grant AST-0926815), the State of Texas, the US Air Force (AFRL FA9451-04-2-0355), and generous support from private individuals and foundations.

Observations were obtained with the Hobby-Eberly Telescope (HET), which is a joint project of the University of Texas at Austin, the Pennsylvania State University, Ludwig-Maximilians-Universit\"at M\"unchen, and Georg-August-Universit\"at G\"ottingen. The HET is named in honor of its principal benefactors, William P. Hobby and Robert E. Eberly.

VIRUS is a joint project of the University of Texas at Austin, Leibniz-Institut f\"ur Astrophysik Potsdam (AIP), Texas A\&M University (TAMU), Max-Planck-Institut f\"ur Extraterrestrische Physik (MPE), Ludwig-Maximilians-Universit\"at Muenchen, Pennsylvania State University, Institut fur Astrophysik G\"ottingen, University of Oxford, and the Max-Planck-Institut f\"ur Astrophysik (MPA). In addition to Institutional support, VIRUS was partially funded by the National Science Foundation, the State of Texas, and generous support from private individuals and foundations.

The authors acknowledge the Texas Advanced Computing Center (TACC) at The University of Texas at Austin for providing high performance computing, visualization, and storage resources that have contributed to the research results reported within this paper. URL:http://www.tacc.utexas.edu

The Institute for Gravitation and the Cosmos is supported by the Eberly College of Science and the Office of the Senior Vice President for Research at the Pennsylvania State University.

KG acknowledges support from NSF-2008793.  EG acknowledges support from NSF grant AST-2206222. ASL acknowledges support from Swiss National Science Foundation. 
HK and SS acknowledges the support for this work from NSF-2219212. EK and SS are supported in part
by World Premier International Research Center Initiative (WPI Initiative), MEXT, Japan.

\facility{HET}

\software{Astropy \citep{astropy:2018}, NumPy \citep{numpy}, SciPy \citep{SciPy}, Matplotlib \citep{matplotlib}, EliXer \citep{Davis_2023}}


\clearpage

\bibliography{lya.bib}


\end{document}